\documentstyle[preprint,eqsecnum,aps]{revtex}

\def\J#1#2#3#4{#1~{\bf #2}, #3 (#4).}
\def\K#1#2#3#4{#1~{\bf #2}, #3 (#4)}

\def\PRB{Phys. Rev. B}
\def\PRA{Phys. Rev. A}

\def\RMP{Rev. Mod. Phys.~}

\def\JPC{J. Phys. Chem.}
\def\JPC2{J. Phys. C: Solid State Phys.}
\def\JCP{J. Chem. Phys.}
\def\CPL{Chem. Phys. Lett.}

\def\N{Nature (London)}
\def\S{Science}
\def\PRL{Phys. Rev. Lett.~}

\def\CP{Chem. Phys.}
\def\JMST{J. Mol. Struct. (THEOCHEM)}

\def\ibid{{\it ibid.}}
\def\ea{{\em et al.}}

\def\mun{$\mu_{\rm N}$}
\def\mun2{$\mu_{\rm N_{2}}$}

\begin{document}
\draft

\title{Topology and Energetics of Metal-Encapsulating Si 
Fullerene-Like Cage Clusters}
\author{Takehide Miyazaki$^{a}$, Hidefumi Hiura$^{b}$ and Toshihiko 
Kanayama$^{a}$
}
\address{
$^{a}$Advanced Semiconductor Research Center,
National Institute of Advanced Industrial Science and 
Technology, AIST Tsukuba Central 4,\\
1-1-1 Higashi, Tsukuba 305-8562, Japan \\
% Phone: +81-298-61-5404, Fax: +81-298-61-2576 \\
% E-mail: takehide.miyazaki@aist.go.jp
}
\address{$^{b}$Fundamental Research Laboratories,
NEC Corporation,\\
34 Miyukigaoka, Tsukuba 305-8501, Japan}
\date{\today}
\maketitle
\begin{abstract}On the basis of 
a topological discussion as well as an {\it ab initio} 
calculation, we show that it is possible 
to construct a fullerene-like Si cage by doping of a transition 
metal atom. The cage is a simple 3-polytope which maximizes the 
number of its inner diagonals close to the metal atom. Our 
topological argument also reveals how closely the structure of the 
fullerene-like Si cages studied is related to that of fullerenes 
themselves.
\end{abstract}
% \twocolumn
\clearpage
\tightenlines
Synthesis of fullerene-cage-like clusters composed of elements 
other than carbon (C) is a subject of great 
interest\cite{Quong}. Especially, it is 
intriguing to study whether silicon (Si) analogues of the fullerenes 
can exist in energetically favorable forms. Production of Si cage 
clusters is also important from a technological point of view. They 
may be used as building blocks for fabrication of various structures 
in electronic devices in ten nanometer scales. 

A main question which we would like to answer in this study is how 
one can obtain a smooth fullerene-like cage composed of Si atoms.
Due to their $sp^{3}$ nature, the Si atoms tend to bind themselves {\it 
against} generation of fullerene cages. As a matter of fact,
Si$_{n}$ clusters ($n$ up to $\sim$50) usually favor compact forms 
which are completely different from fullerene cages\cite{Honea,Ho}. 
The structure of a Si$_{60}$ cluster in a fullerene cage is highly 
distorted\cite{Li}. 

As is well known, the structure of a typical fullerene cage (C$_{n}$)
is composed of only 5- and 6-membered rings. 
The number of the 5-membered rings is 
twelve irrespective of $n$. Each atom is connected 
with three neighbors. Stability of fullerenes may be attributed to two 
factors: $\sigma$ bonding among $sp_{\|}^2$ hybrids, where $p_{\|}$ 
stands for the intra-cage component of the C $p$ orbital, and $\pi$ 
conjugation among $p$-orbital components normal to the cage surface 
($p_{\perp}$). 
 
A fundamental difficulty to maintain a smooth Si fullerene cage stems 
from the fact that the $\pi$ conjugation among the $p_{\perp}$ orbitals 
does not occur. Thus, the caging mechanism is only the $\sigma$ bonding 
of the $sp_{\|}^2$ hybrids. The distortion of the fullerene cage of 
Si$_{60}$ is due to admixture of substantial $p_{\perp}$ components 
with $sp_{\|}^2$.

A promising solution to the augmentation of Si cage structures is 
to put one or more additional atoms inside of the cages. If electron 
orbitals of these ``doped'' atoms have a substantial overlap with the 
$p_{\perp}$ orbitals, then the additional cohesion forces would be 
supplied to the cage. This idea has 
originally sprouted from the construction of Si$_{n}$ clusters with 
$n$ larger than $\sim$20\cite{Kaxiras} to account 
for the exceptional chemical inertness of Si$_{39}$ and 
Si$_{45}$\cite{Elkind} and also for ``prolate-oblate'' structural 
change in Si$_{n}$ around 
$n$$\sim$27\cite{Jarrold1}. A common aspect
of the Si$_{n}$ configurations studied in 
Refs.\cite{Kaxiras} is that they are configured to 
mimic as much of the bulk Si structure as possible, while the 
outermost cages yet resemble the counterparts of fullerenes.

As explained above, there are at least two mechanisms to stabilize a 
doped Si fullerene-cage cluster: 
(a) $\sigma$ bonding within the $sp_{\|}^{2}$ hybrid network 
of the cage and (b) bonding among the $p_{\perp}$ orbitals of 
the cage and the orbitals of doped atoms. A key issue to be solved for 
production of stable fullerene-like Si cage clusters is how these two 
factors should be 
tuned in order to maximize the total binding among the constituent 
atoms in the clusters. 

The goal of this study is to give an explicit answer to this question 
from a theoretical point of view. For this purpose, we target metal-doped 
Si cage clusters, $M{@}$Si$_{n}$, because we have recently experimentally 
suggested that stable metal-encapsulating Si cage clusters 
($M{@}$Si$_{n}$, $M$$=$Hf, Ta, W, Re, Ir, etc. and $n$$\leq$$\sim$14) 
might be grown\cite{Hiura}. We also note the outstanding property 
of $M{@}$Si$_{n}$. When doped with Na and Ba, the $M{@}$Si$_{n}$ 
clusters in certain phases of the solid state  
exhibit superconductivity\cite{Kawaji}. Although some 
theoretical\cite{Wei} as well as experimental\cite{Beck}
studies have been performed for metal-doped Si clusters, roles of the 
cage topologies played in stabilization of fullerene-like Si cages 
have been little argued.

First, we discuss what topology should be suitable for a cage of 
$M{@}$Si$_{n}$ ($n$$\leq$20). A very important clue 
to the solution has been recently given by mathematicians, 
Bremner and Klee\cite{Bremner}. They study inner diagonals
(see below or Ref.\cite{Bremner} for definition of 
an inner diagonal) of convex polytopes. The quantity of our own 
interest here
is the number of inner diagonals denoted by $\delta_{3}$.
We will propose that a simple 3-polytope\cite{Ziegler}, in which the 
number of inner diagonals close to the dopant ($\delta_{3}^{eff}$) is 
maximized, should be a candidate of a stable fullerene-like
Si cage of $M{@}$Si$_{n}$. 

Next, we perform a first-principles energetics of the clusters
in the case of $M$$=$W (W${@}$Si$_{n}$, 8$\leq$$n$$\leq$16). 
The clusters in fullerene-like cages predicted by the 
above topological picture are indeed obtained for $n$$=$12 [Fig.1(d)] 
and $n$$=$14 [Fig.1(e)]. For $n$$=$8 and 10, however, the caging with 
corresponding simple 3-polytopes [Fig.1(a) and Fig.1(b)] may not occur, 
because their total energies are significantly higher than those of 
the respective isolated atoms. For $n$$=$16, a single cage is unstable 
(Fig.2). The cage is relaxed 
into a skeleton similar to either of $n$$=12$ [Fig.1(d)] or $n$$=$14 
[Fig.1(e)]. Thus we conclude that it is possible to construct 
energetically favorable fullerene-like Si cage clusters doped with a 
transition metal atom, whose cage sizes depend on the dopant.

Now we describe our results. First a topological discussion of Si cages
is given. We denote the number of 
$p$-membered rings of a Si cage as $N_{p}$.
Throughout this study, we limit the range of $p$ to 4$\leq$$p$$\leq$6.
We also use a vector notation 
$\vec{N}$$=$$(N_{4},N_{5},N_{6})$ for brevity. 
For example, simple 3-polytopes with $n$$=$8 and 10 vertices are
a cube [$\vec{N}$$=$(6,0,0), Fig.1(a)\cite{iwk}] and a pentagonal prism 
[$\vec{N}$$=$(5,2,0), Fig.1(b)], respectively. For each $n$$\geq$12, 
simple 3-polytopes with different 
$\vec{N}$'s exist as shown in Fig.1 and Table I\cite{preliminary}. 

An inner diagonal (or a 3-diagonal) of a convex 3-polytope $P$  
is defined as a segment that joins two vertices of $P$ and that lies, 
except for its ends, in $P$'s relative interior\cite{Bremner}. 
There are two other kinds of diagonals of $P$, 1- and 2-diagonals. 
The former are 
the edges. As for the latter, there are two, five and nine 
2-diagonals in a 4-, 5- and 6-membered ring, respectively. If we 
denote the number of $i$-diagonals by $\delta_{i}$, then 
$\delta_{1}$$+$$\delta_{2}$$+$$\delta_{3}$$=$$\frac{n(n-1)}{2}$, 
where $n$ is the number of vertices.
For example, $\delta_{3}$$=$4 for $\vec{N}$$=$(6,0,0) and 
$\delta_{3}$$=$10 for $\vec{N}$$=$(5,2,0). In this study,
we further define $\delta_{3}^{eff}$ as 
$\delta_{3}^{eff}$$=$$\sum_{i>j}\theta(d_{\rm cut}-d_{ij})$, where 
$i$ and $j$ are indices of Si atoms, $d_{\rm cut}$ is the cut-off 
radius and $d_{ij}$ is the distance between a metal atom and a diagonal 
joining the $i$-th and $j$-th Si atoms. $\theta(x)$ is a step function;
$\theta(x)$$=$1 for $x$$\geq$0 and $\theta(x)$$=$0 otherwise.
Obviously, $\delta_{3}^{eff}$ approaches $\frac{n(n-1)}{2}$ for 
$d_{\rm cut}$$\rightarrow$$\infty$.
In order to count the number of only inner diagonals, $d_{\rm cut}$ 
must be small. In practice, we vary $d_{\rm 
cut}$ from 1.0{\AA} down to 0.25{\AA} to check the $d_{\rm 
cut}$-dependence on $\delta_{3}^{eff}$.
The overlap between the $p_{\perp}$ orbitals of a Si cage and
the counterparts of the dopant metal atom may be roughly 
proportional to the number of the $p_{\perp}$ orbitals 
pointing toward the metal atom, which should also be approximately 
proportional to $\delta_{3}^{eff}$ at a small $d_{\rm cut}$ by its definition.
Thus analyzing $\delta_{3}^{eff}$ may measure the strength of the 
bonding
between the doped atom and the Si cage. Our first-principles 
calculation shows that $\delta_{3}^{eff}$ at $\delta_{\rm cut}$$=$0.25{\AA} 
is maximized for the lowest-energy W${@}$Si$_{n}$ cluster 
for each of $n$$=$12 and 14.

Bremner and Klee\cite{Bremner} show that, for a given $n$, $P$ 
with a maximum $\delta_{3}$ is {\it simplicial}\cite{Ziegler}. 
A simplicial 3-polytope seems the best 
for a cage of $M{@}$Si$_{n}$ for each $n$, because the energy gain 
due to the factor (b) should be fully enjoyed. This is not true, 
however, 
since the mechanism (a) is instead sacrificed so that a {\it total} 
energy gain is not maximized. Our first-principles calculation
shows that a W${@}$Si$_{12}$ cluster with an 
icosahedron cage ({\it i.e.} a simplicial 3-polytope with 12 vertices
and $\delta_{3}$$=$36) is energetically very unfavorable [5.76 eV 
higher in energy than the (6,0,2)-cage cluster shown in FIg.1(d)],
despite that {\it all} $p_{\perp}$ orbitals of the cage point toward 
the central W atom. The averaged Si-Si distance ($d_{2}$) of the 
icosahedron cage (2.72{\AA}) is much longer than those of both 
(4,4,0)- ($d_{2}$$=$2.42{\AA}) and (6,0,2)-cages 
($d_{2}$$=$2.39{\AA}). In contrast, the averaged Si-W distances 
($d_{1}$) are similar, $\sim$2.6{\AA}. This evidences that too much 
preference of the factor (b) is taken over (a) in the icosahedron cage.
Therefore a {\it simple} 
3-polytope is a good candidate for a Si$_{n}$ ($n$$\leq$20) cage, 
in which
both (a) and (b) factors may work equally well to enhance the 
stability of the $M{@}$Si$_n$ cluster as a whole. 

Turning to the topology of a {\it carbon} fullerene cage, 
its peculiarity can be characterized in light of $\delta_{3}$. 
Let $Q$ be a simple 3-polytope with $n$ vertices.
For each of $n$$\geq$24, $Q$ with $F$ ($=n/2+2$) facets
composed of only twelve pentagons and $F$$-$12 
hexagons maximizes $\delta_{3}$ which is $n(n-13)/2+30$\cite{GM}. 
For $n$$=$20, $Q$ with the largest $\delta_{3}$ ($=$100) is a 
dodecahedron which has only twelve pentagons. In the case of 
$n$$=$22, however, $Q$ with $N_{5}$$=$12 cannot be 
realized\cite{Bremner}. The maximum possible value of $\delta_{3}$ is 
128 when $\vec{N}$$=$(1,10,2).
Thus a typical fullerene C$_{n}$ cage ($n$$\geq$20)
can be identified to be a simple 3-polytope with a maximum value of 
$\delta_{3}$ except $n$$=$22. The Si$_{n}$ cages we consider for 
8$\leq$$n$$\leq$16 include $Q$'s with maximum $\delta_{3}$'s. Although
$\delta_{3}$ ($=$18) of the (6,0,2)-cage [Fig.1(d)] is not a maximum, 
it is the second largest value. This is why we can say that our Si 
cages are fullerene-like.

In order to substantiate the above topological argument, we perform a 
first-principles energetics of the W${@}$Si$_{n}$ clusters. Here we 
adopt a recently proposed single-parent evolution algorithm 
(SPEA)\cite{SPEA} to update the atomic coordinates in an unbiased way.
For SPEA simulation, we calculate the total energies of clusters 
with a linear combination of atomic orbitals (LCAO) using
the GAUSSIAN98 package\cite{G98}.
Then we use the calculated coordinates of the clusters as the inputs of 
plane-wave (PW\cite{convergence}) total energy calculations\cite{STATE} 
where quenched molecular dynamics is performed for final convergence.
Electronic structures of clusters are 
calculated with density-functional theory\cite{DFT} within 
generalized gradient approximation to the exchange-correlation energy
functionals (Becke'88\cite{B88} and Perdew-Wang'91\cite{PW91} for LCAO 
and Perdew-Wang-Ernzerhof'96\cite{PBE96} for PW)\cite{ECPPP}. 

With the aid of SPEA, we indeed find that the cages of energetically 
favorable 
W${@}$Si$_{12}$ and W${@}$Si$_{14}$ clusters are simple 3-polytopes 
with $\vec{N}$$=$(6,0,2) [Fig.1(d)] and (3,6,0) [Fig.1(e)], respectively. 
Varying $d_{\rm cut}$ from 1.0{\AA} down to 0.25{\AA}, we 
find that $\delta_{3}^{eff}$'s of both are constants (6 and 4; see 
Table I for the summary of the result).
On the other hand, $\delta_{3}^{eff}$'s of the other W${@}$Si$_{12}$ 
and W${@}$Si$_{14}$ clusters decrease to 
0$-$2 as $d_{\rm cut}$ becomes smaller, whose total energies are all 
higher than those of the (6,0,2) and (3,6,0) cages, respectively. 
These results support the topological argument made above. 

It should be noted that the W${@}$Si$_{n}$ ($n$$\sim$10) 
clusters may be viewed as 
 metastable phases of Si-rich tungsten silicides, which do not appear 
in equilibrium W-Si phase diagram\cite{Gmelin}. It is suggested in the
previous experiment\cite{Hiura} that the clusters are obtained via a 
sequential growth where a W${@}$Si$_{n}$ cluster is obtained by 
attaching additional $m$ Si atoms to a smaller W${@}$Si$_{n-m}$ cluster.
In order to understand the essence of the growth process on the 
basis of our energetics, 
we calculate differential binding energy, defined as 
$\Delta E(n)$$=$$E_{\rm 
bind}(n+2)-2E_{\rm bind}(n)+E_{\rm bind}(n-2)$, where
$E_{\rm bind}(n)$$=$
$E_{\rm total}^{atom}({\rm W})+nE_{\rm total}^{atom}({\rm Si})
-E_{\rm total}^{cluster}({\rm WSi}_{n})$. 
A negative value of $\Delta E(n)$ means that generation of 
a W${@}$Si$_{n}$ cluster is favorable. 
We find that $\Delta E(10)$$=$1.6 eV, $\Delta E(12)$$=$$-$2.8 eV and
$\Delta E(14)$$=$$-$0.3 eV. For $n$$=$10, stable caging of a W atom 
in simple 3-polytopes [Fig.1(b)] is unlikely to occur. 
The cluster should be chemically very reactive, because there 
are too few Si atoms around W to fulfil its reaction points.
If additional Si atoms can arrive at the 
cluster prior to its collapse, 
then subsequent growths to W${@}$Si$_{12}$ and possibly to W${@}$Si$_{14}$
should occur with high probabilities.

We also find that there exists a threshold cage size for 
W${@}$Si$_{n}$ beyond which the cages become unstable. In Fig.2, 
some of calculated structures of W${@}$Si$_{16}$ clusters are 
shown. The initial configurations of the cages are taken like simple 
3-polytopes with 16 vertices. Upon relaxation, all cages we considered 
are completely distorted and become similar to either 
the (6,0,2) cage plus 2 Si atoms or
the (3,6,0) cage plus 4 Si atoms. It appears that $n$$=$14 is the 
threshold size in the case of W doping. For a cage beyond a critical 
size, there are too many Si atoms around W to cover its reaction 
points. 

In conclusion, we show that it is possible to construct Si clusters in 
fullerene-like cages with transition-metal atom doping. A topological 
discussion suggests that simple 3-polytopes with maximum numbers of 
inner diagonals close to the dopant may be good candidates of 
fullerene-like Si cages. First-principles calculation shows that, in 
the case of W doping, W${@}$Si$_{12}$ and W${@}$Si$_{14}$ are 
energetically most favorable and also have the cages predicted by 
the above topological picture.

This work is partly supported by NEDO.

\narrowtext

\begin{center}
\begin{table}
     \caption{Topological properties and energetics of W${@}$Si$_{n}$
     clusters. For definitions of $n$, $\vec{N}$, $\delta_{3}$, 
     $\delta_{3}^{eff}$ and $d_{\rm cut}$, see text. $d_{1}$ and 
     $d_{2}$ are the Si-W and Si-Si distances in {\AA}, averaged 
     over the respective pairs whose separations are less than 2.7{\AA}.
     $\Delta E$ is the total energy ( eV) relative to that of the 
     lowest-energy cluster, $\vec{N}$$=$(6,0,2) for $n$$=$12 and 
     $\vec{N}$$=$(3,6,0) for $n$$=$14. }
 \begin{tabular}{ccccccccccccc}
     $n$ & $\vec{N}$ & $\delta_{3}$ & \multicolumn{5}{c}
     {$\delta_{3}^{eff}$($d_{\rm cut}$)} &  $d_{1}$ & 
     $d_{2}$& structure & $\Delta E$\\
       &   &   & $d_{\rm cut}$({\AA})$=$&1.0 & 0.6 & 0.3 & 0.25 &  
       &   \\
     \hline 
     12 & (4,4,0) & 20 && 8 & 4 & 0 & 0 & 2.59 & 2.42 
     &Fig.1 (c) & 2.16\\
     12 & (6,0,2) & 18 && 6 & 6
     & 6 & 6 & 2.67 & 2.39 & Fig.1 (d) & 
     0.00\\
     \hline
     14 & (3,6,0) & 34 && 4 & 4 & 4 & 4 & 2.69 & 2.34
     &Fig.1 (e) & 0.00 \\
     14 & (4,4,1) & 33 && 11& 5 & 1 & 0 & 2.66 & 2.35 
     &Fig.1 (f) & 0.95 \\
     14 & (5,2,2) & 32 && 13& 4 & 2 & 2 & 2.64 & 2.39 
     &Fig.1 (g) & 0.63 \\
     14 & (6,0,3) & 31 && 9 & 7 & 4 & 0 & 2.63 & 2.39 
     &Fig.1 (h) & 0.89
\end{tabular}
\end{table}
\end{center}

\clearpage

\begin{center}
        {\large Figure captions}
\end{center}

\begin{itemize}
\item[Figure 1] (color) Calculated structures of W${@}$Si$_{n}$ in 
``fullerene-like'' cages. Double line segments  
are drawn when the inter-atom distance is less than 2.7{\AA}. The 
double line segments corresponding to the edges of simple 
3-polytopes are colored with blue. In panels (a), (g) and (h), 
 red single lines should be supplemented and also green double lines 
 should be removed to 
retrieve connectivity in the corresponding simple 
3-polytopes. See also a note\cite{iwk}.
\item[Figure 2] (color) Calculated structures of W${@}$Si$_{16}$. Si 
atoms colored with dark blue are regarded ``attached'' to either 
the (6,0,2)-cage W${@}$Si$_{12}$ (panel (e)) or 
(3,6,0)-cage W${@}$Si$_{14}$ 
clusters (panels (a), (b), (c) and (d)). 
Other conventions are the same as in Fig.1.
\end{itemize}


\begin{references}
\bibitem{preliminary} Suppose that $P$ has $E$ edges and $F$ facets, then 
$E$$=$$(4N_{4}+5N_{5}+6N_{6})/2$ and $F$$=$$N_{4}+N_{5}+N_{6}$.
Since $P$ is simple, $E$$=$$3n/2$, meaning that $n$ must be even. 
Combining these with the Euler's 
formula $n$$+$$F$$=$$E$$+$2, we get $2N_{4}+N_{5}$$=$12 and 
$N_{5}+2N_{6}$$=$$n$$-$8. Note that $P$'s with $\vec{N}$$=$(6,0,1) 
and (5,2,1) do not exist. The two conditions obtained here are necessary 
but not sufficient for the existence of $P$.
\bibitem{convergence} 
Each cluster is located in a simple cubic supercell with the edge 
length of 30 a.u. in PW calculation. The self-consistent field iterations are continued
until the total energy converges to $\sim$3$\times$10$^{-6}$eV / atom.
Only a $\Gamma$ point is sampled for Brillouin-zone integration.
The atomic positions of all atoms in supercell are relaxed
until the residual forces are less than $\sim$0.06 eV/{\AA}.
\bibitem{ECPPP} 
Coulomb interactions between electrons and 
ions are approximated with effective-core potentials\cite{ECP} (for LCAO) and
pseudopotentials (PP's; for PW). In PW calculation, PP's of W and Si 
are as follows; For W, 5$p$, 5$d$ 
and 6$s$ states are considered, where 5$p$ and 5$d$ orbitals are 
pseudized with ultra-soft PP's\cite{USPP} and 6$s$ with a 
norm-conserving PP (NCPP)\cite{NCPP}. Only NCPP's are 
constructed for 3$s$ and 3$p$ states of Si.
\end{references}
\end{document}